\documentstyle[12pt,epsf]{article}

\newcommand{\be}{\begin{equation}}
\newcommand{\ee}{\end{equation}}
\newcommand{\ba}{\begin{eqnarray}}
\newcommand{\ea}{\end{eqnarray}}
\newcommand{\bb}{\begin{thebibliography}}
\newcommand{\eb}{\end{thebibliography}}
\newcommand{\ci}[1]{\cite{#1}}
\newcommand{\bi}[1]{\bibitem{#1}}
\newcommand{\lab}[1]{\label{#1}}
\begin{document}

\begin{center}
{\bf LARGE - DISTANCE EFFECTS IN QCD AND SPIN ASYMMETRIES IN DIFFRACTIVE REACTIONS.}

\vskip 5mm
S.V.Goloskokov$^{\dag}$

\vskip 5mm

{\small
{\it
Bogoliubov Laboratory of Theoretical  Physics,\\
 Joint Institute for Nuclear Research,\\
Dubna 141980, Moscow region, Russia
}
\\
$\dag$ {\it
E-mail: goloskkv@thsun1.jinr.ru
}}
\end{center}

\vskip 5mm

\begin{center}
\begin{minipage}{150mm}
\centerline{\bf Abstract}
Large - distance effects in the hadron wave function for the 
hadron-Pomeron vertex in the QCD -model are discussed. It is 
shown that they lead to the spin-flip part of the Pomeron 
coupling which can reach $10 \div 20 \%$ with respect to the 
non-flip part. The obtained structure of the Pomeron
coupling  modifies  spin asymmetries in high--energy 
diffractive reactions where  the Pomeron structure becomes 
determinative. We study spin asymmetries in exclusive reactions 
and in the diffractive $Q \bar Q$ and $J/\Psi$ production. We have 
found that  spin effects in these diffractive processes are  
sensitive to the spin-flip part of the Pomeron coupling.
\\
{\bf Key-words:}
Large - distance effects, polarization, diffractive reactions,
Pomeron, spin asymmetries
\end{minipage}
\end{center}
\vskip 10mm

Study of diffractive processes at HERA provides an excellent 
possibility of investigating the nature of the Pomeron. New 
results on the Pomeron intercept in diffractive events, 
information about the Pomeron partonic structure,  have been 
obtained in H1 and ZEUS experiments \ci{h1_zeus}. Investigation 
of the diffractive  vector meson and $Q \bar Q$ production should 
play a keystone role in future study of this object.  Really, 
diffractive reactions give an important information on the 
Pomeron structure which has a nonperturbative character. At the 
same time, they can be used to study the gluon distribution in 
the nucleon at small $x$ \ci{rys}. 

The Pomeron is a color singlet exchange which describes  high 
energy reactions at fixed momentum transfer. The two-particle 
amplitude determined by the Pomeron exchange can be written in 
the form
\be
\lab{tpom}
\hat T(s,t)=i I\hspace{-1.6mm}P(s,t) V_{h_1h_1I\hspace{-1.1mm}P}(p_1,t)
 \otimes
V^{h_2h_2 I\hspace{-1.1mm}P}(p_2,t).    
\ee

Here  $V_{\mu}^{hhI\hspace{-1.1mm}P}$ are the Pomeron-hadron 
vertices of the particles with initial momenta equal to $p_1$ and 
$p_2$, and $I\hspace{-1.6mm}P$ is a function caused by the Pomeron. 
The calculation of this amplitude in the nonperturbative 
two-gluon exchange model  \ci{la-na} and in the BFKL model 
\ci{bfkl} shows that the Pomeron couplings are simple in form 
(the standard coupling in what follows):
\be 
\lab{pmu} 
V^{\mu}_{hh I\hspace{-1.1mm}P} =B_{hh 
I\hspace{-1.1mm}P}(t)\; \gamma^{\mu}.  
\ee 
In this case the spin-flip effects are suppressed as a power of 
$s$.

Since the Pomeron consists of a two-gluon \cite{low}, the Pomeron coupling should have two gluon indices. The Pomeron
 coupling with the proton can be written in the form
\begin{equation}
\label{ver}
V_{pgg}^{\alpha\alpha'}(p,t,x_P)= 4 p^{\alpha} p^{\alpha'}
A(t,x_P)+(\gamma^{\alpha} p^{\alpha'} +
\gamma^{\alpha'} p^{\alpha}) B(t,x_P).
\end{equation}
where $r$ is the momentum transfer and 
$x_P$ is the fraction of the initial proton momentum carried by 
the gluon system. 

Let us consider the single-flip amplitude of the elastic $pp$ 
scattering (\ref{tpom}) with the vertex (\ref{ver}). In this 
case, we can fix the helicities of the one proton line at $+1/2$. 
The  amplitude is then simplified to
\be
\label{t}
T_{ \lambda_4  +; \lambda_2 +} = F_{ \lambda_4  \lambda_2} =
\bar u(p_4,\lambda_4) \bar u(p_3,+)\hat T(s,t)
u(p_2,\lambda_2) u(p_1,+).
\ee
The leading term of the non-flip matrix element of the 
coupling (\ref{ver}) is found 
to be  $\bar u(p_3,+)V_{pgg}^{\alpha\alpha'}(p_1) u(p_1,+) 
\propto p_1^{\alpha} p_1^{\alpha'}$. The $1/s$ term appears from 
the $\delta$-function integration in the gluon loop. Finally, the $F_{ \lambda_4  \lambda_2}$  amplitude can be written 
as follows:
\be
F_{ \lambda_4  \lambda_2} =
\bar u(p_4,\lambda_4)[s A(t) +
/\hspace*{-0.20cm} p_1 B(t)] u(p_2,\lambda_2) \phi(t).
\ee
Here  $\phi(t)$  represents an additional function of $t$ from  
the  spin-independent Pomeron coupling and the Pomeron exchange.

The proton--proton helicity-non-flip and helicity-flip amplitudes 
are expressed in terms of the $A(t)$ and $B(t)$ functions 
\begin{equation}
\label{fnf}
F_{++}(s,t)= i s [B(t)+ 2 m A(t)] \phi(t);\;\;
F_{+-}(s,t)= i s \sqrt{|t|} A(t) \phi(t),
\end{equation}
where $m$ is a proton mass.

Thus, $(\gamma^{\alpha} p^{\alpha'} + \gamma^{\alpha'} 
p^{\alpha}) B(t)$ in (\ref{ver}) is a standard Pomeron coupling 
like  (\ref{pmu}) which determines the spin-non-flip amplitude. 
The term $ p_{\alpha}p_{\alpha'} A(r)$ leads to the spin-flip in 
the Pomeron vertex which does not vanish in the $s \to \infty$ 
limit. Note that in the model \cite{lansh-m}, the Pomeron 
effectively couples to the hadron like a $C= +1$ isoscalar 
photon. Then, the vertex (\ref{ver}) is equivalent to the 
isoscalar electromagnetic  nucleon current with the Dirac and 
Pauli nucleon form factors \cite{nach}. A similar form of the 
proton-Pomeron coupling has been used in \cite{schaefer}. In the 
model \cite{gol_mod} the form (\ref{ver}) was found to be valid 
for small momentum transfer $|t| <  \mbox{few GeV}^2$ and the 
$A(t)$ amplitude was caused  by the meson-cloud effects in the 
nucleon. In a QCD--based diquark model of the proton 
\cite{kroll},  the $pgg$ coupling in the form (\ref{ver}) has 
been found at moderate momentum transfer \cite{gol_kr}. There, 
the $A(t)$ contribution is determined by the effects of vector 
diquarks inside the proton, which are of the order of $\alpha_s$. 
In all the cases the spin-flip $A(t)$ contribution is  determined  
by the nonperturbative effects in the proton. 

The absolute value of the ratio of $A$ to $B$ is proportional to 
the ratio of helicity-flip and non-flip amplitudes. This ratio  
is about $|A|/|B| \sim 0.1 -0.2 \,\mbox{GeV}^{-1}$ 
\cite{gol_mod,gol_kr} and has a weak energy dependence. In the 
models \cite{gol_mod,gol_kr}, the spin-flip amplitude is not in 
the phase with the non-flip amplitude. As a result, the 
single-spin asymmetry appears 
\be 
\lab{epol} A_N= -2 \frac{{\rm 
Im}[F_{++} F_{+-}^*]}{|F_{++}|^2+2 |F_{+-}|^2}. 
\ee
 
In the model \ci{gol_mod}, the amplitudes $A$ and $B$ have a phase 
shift caused by the soft Pomeron rescattering effect. The 
predicted asymmetry \cite{akch} is shown in  Fig.\ 1.  The 
diquark model \cite{gol_kr} results in the helicity flips which 
are generated by vector diquarks. The $A$ amplitude is out of 
phase with the Pomeron contribution to the amplitude $B$ too. The 
model provides a single-spin asymmetry $A_N$ shown in Fig.\ 2 
which is rather large for momentum transferred $|t| \ge 3\, 
\mbox{GeV}^2$. 

\phantom{.}
\vspace*{-1.9cm}
\begin{minipage}{8.5cm}
\phantom{.}
\hspace{1.5cm}
\vspace{.5cm}
\epsfxsize=7cm
\epsfbox{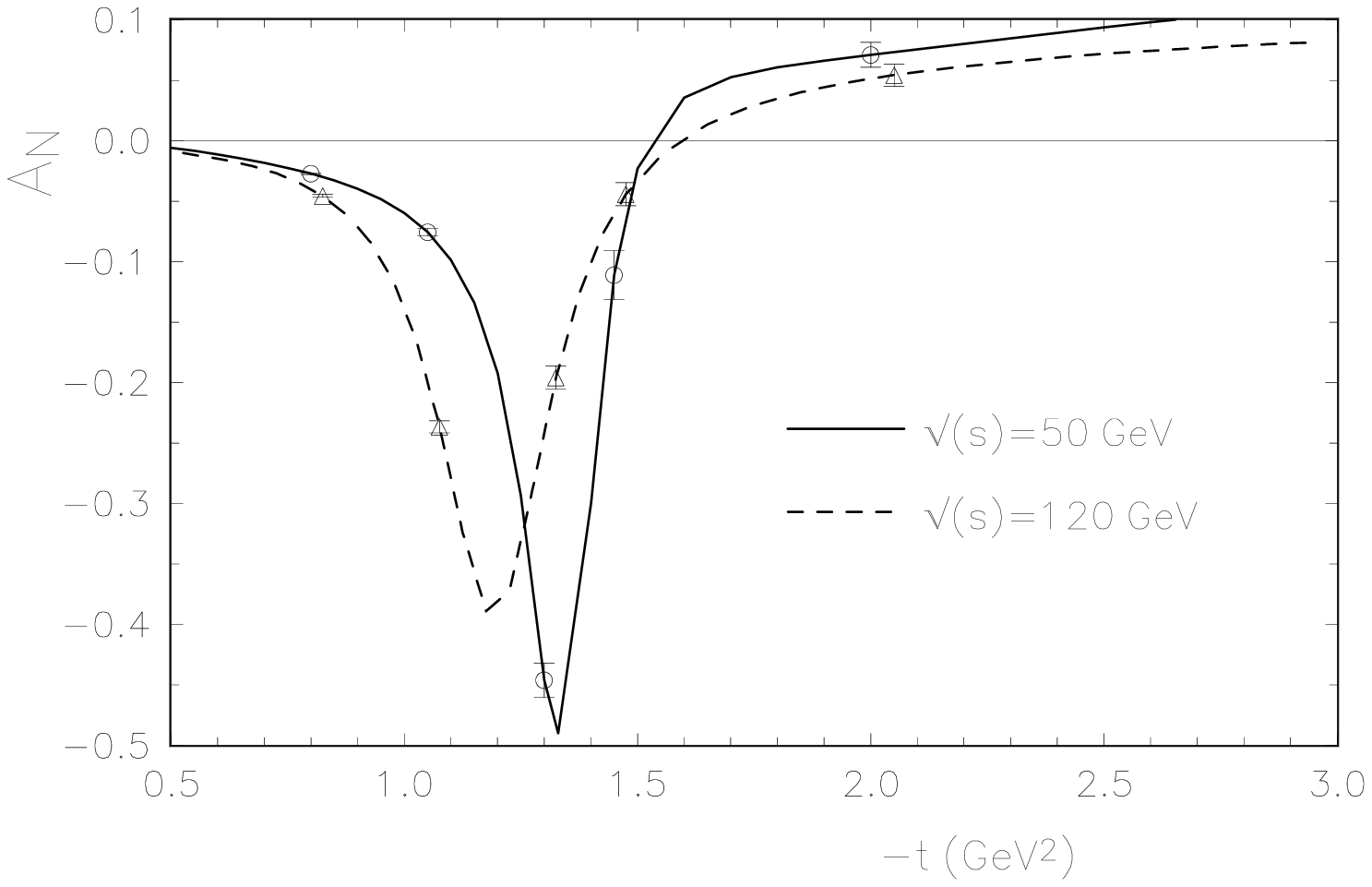}
\end{minipage}
\hspace{-1.cm}
\begin{minipage}{7cm}
Fig.1~ Meson cloud model predictions for single-spin 
transverse asymmetry of the $pp$ scattering at RHIC energies
  \cite{akch}. Error bar indicates expected
 statistical errors for the PP2PP experiment at RHIC.
\end{minipage}
\phantom{.}
\begin{minipage}{8.5cm}
\hspace{.5cm}
\epsfxsize=7.0cm
\epsfbox{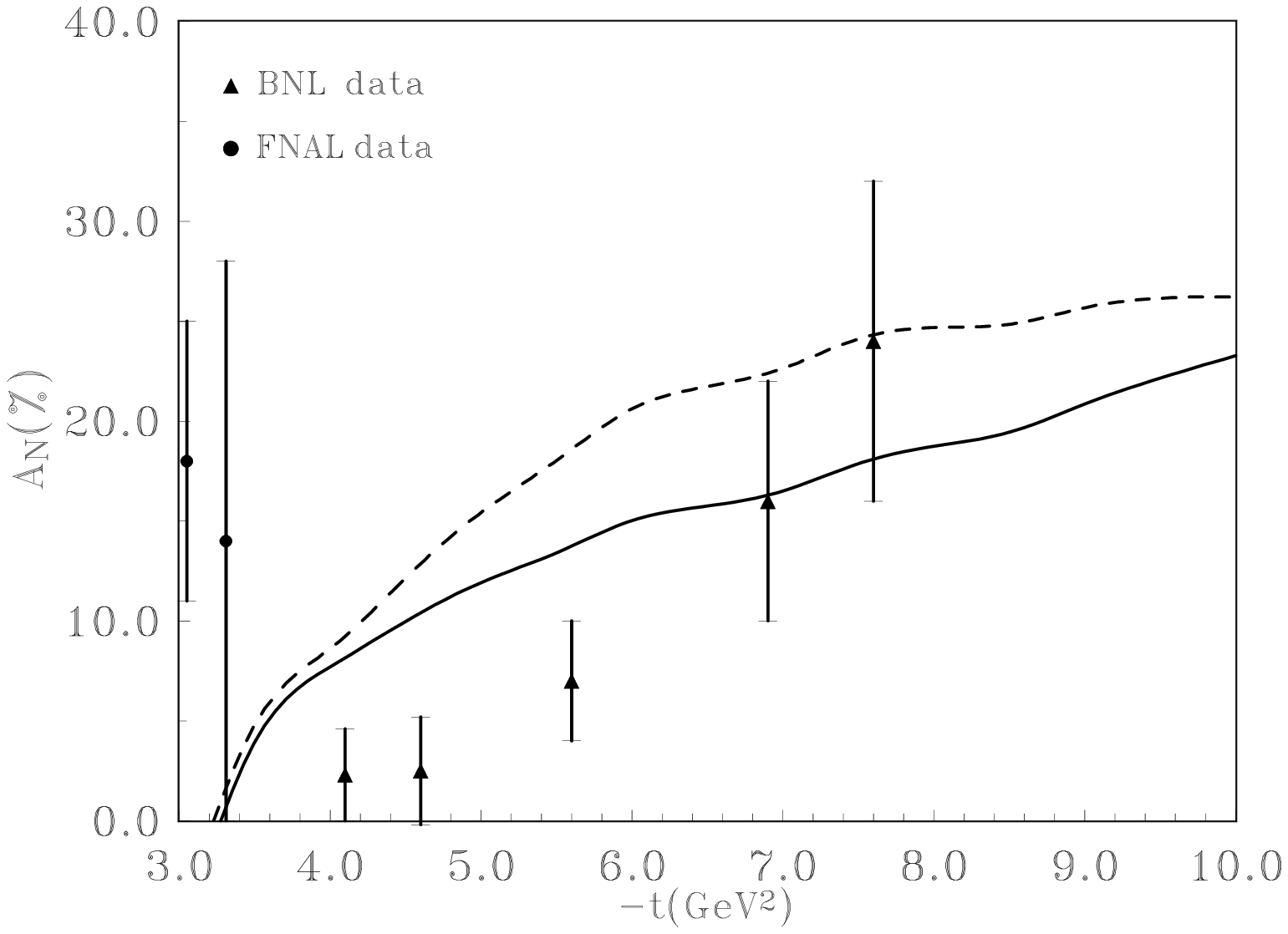}
\end{minipage}
\begin{minipage}{7cm}
Fig.2~  Model predictions for single-spin asymmetry
 for two fits of the $B$ amplitude (see \cite{gol_kr} for details). Data from  G. Fidecaro et al.,
                Phys. Lett. {\bf B105}, 309 (1981) (FNAL) and R.C.\ Fernow, A.D. Krisch,
                Ann. Rev. Nucl. {\bf 31}, 107 (1981) (BNL).
\end{minipage}

A convenient tool to study the spin-dependent Pomeron structure 
might be polarized diffractive leptoproduction reactions. We 
shall consider here the $A_{ll}$ asymmetry of the $J/\Psi$ and $Q 
\bar Q$ production. The cross section of these 
reactions has the following important parts: leptonic and 
hadronic tensors and the amplitude of the $\gamma^\star 
I\hspace{-1.7mm}P \to J/\Psi (Q \bar Q)$ transition. The 
structure of leptonic tensor is quite simple. The hadronic tensor 
for the vertex (\ref{ver}) has the form
\be
\label{wtenz}
W^{\alpha\alpha';\beta\beta'}(s_p)= \sum_{final spin} \bar u(p')    V_{pgg}^{\alpha\alpha'}(p,t,x_P) u(p,s_p) \bar u(p,s_p)
V_{pgg}^{\star\,\beta\beta'}(p,t,x_P) u(p').
\ee
Here $p$ and $p'$ are the initial and final proton momenta, and $s_p$ 
is a spin of the initial proton. 

The spin-average and spin dependent 
cross sections with parallel and antiparallel longitudinal polarization of a lepton and a proton are determined by the relation
\be
\sigma(\pm) =\frac{1}{2} \left( \sigma(^{\rightarrow} _{\Leftarrow}) \pm \sigma(^{\rightarrow} _{\Rightarrow})\right).
\ee
These cross sections can be expressed in terms of 
spin-average and spin dependent value of the lepton and hadron 
tensors. For the latter one can write
\begin{equation}
W^{\alpha\alpha';\beta\beta'}(\pm)=\frac{1}{2}( W^{\alpha\alpha';\beta\beta'}(+\frac{1}{2}) \pm W^{\alpha\alpha';\beta\beta'}(-\frac{1}{2})),
\end{equation}
where $W(\pm\frac{1}{2})$ are the hadron tensors with the  
helicity of the initial proton equal to $\pm 1/2$. The leading term 
of the spin average hadron tensor looks like
\begin{equation}
W^{\alpha\alpha';\beta\beta'}(+) = 16 p^{\alpha} p^{\alpha'} p^{\beta}  p^{\beta'} ( |B(t)+2 m A(t)|^2 + |t| |A(t)|^2).
\end{equation}
It is proportional to the proton-proton cross section up to a  function of $t$.

The  spin-dependent hadron tensor is quite complicated.
It can be represented as a sum of structures which have different nature
\begin{equation}
\lab{w-}
W^{\alpha\alpha';\beta\beta'}(-) = \Delta A_{full}^{\alpha\alpha';\beta\beta'} + \Delta A_1^{^\star \alpha\alpha'}
S_2^{\beta\beta'}- \Delta A_1^{\beta\beta'}
S_2^{^\star \alpha\alpha'}.
\end{equation}
The typical contribution to 
$A_{full}^{\alpha\alpha';\beta\beta'}$ looks like as $2 i |B(t)|^2 
p^{\alpha'} p^{\beta'} \epsilon^{\alpha\beta\gamma\delta} 
(p'-p)_\gamma (s_p)_\delta$. This term has  indices of 
different Pomeron couplings in the $\epsilon$ function. This 
contribution is equivalent to the spin-dependent part which 
can be obtained from the coupling (\ref{pmu}) and called by us  
a full block asymmetry. The other terms have a form of a product 
of the asymmetric part of one proton vertex 
(\ref{ver}) $\Delta A_1^{^\star \alpha\alpha'} \sim [2 i 
A^\star(t)/m p^{\alpha'} \epsilon^{\alpha\gamma\delta\rho} 
p_{\gamma} (p'-p)_{\delta} (s_p)_{\rho}]$ to the symmetric part 
of the other $S_2^{\beta\beta'} \sim [4 B(t)p^{\beta} p^{\beta'} 
]$.

A simple model is considered for the amplitude of the $\gamma^\star 
\to J/\Psi$ transition.  The virtual photon is going to the $q 
\bar q$ state and the $q \bar q \to V$ amplitude  is described by 
a non-relativistic wave function \cite{rys,diehl} which is 
regarded as a $S$-wave system of $c \bar c$ quarks. In 
this approximation, quarks have the same momenta equal to half of 
the vector meson momentum. The $J/\Psi$-wave function has the form 
$g (\hat k+m_c) \gamma_\mu$ where $k$ is the momentum of $c \bar 
c$- quarks and $m_c=m_J/2$. The coupling constant $g$ can be 
expressed through the $e^+ e^-$ decay width of the $J/\Psi$ meson. 
The gluons from the Pomeron are coupled with the single and 
different quarks in the $c \bar c$ loop. This ensure the gauge 
invariance of the final result.

The cross section of the $J/\Psi$ leptoproduction can be written in
the form
\be 
\label{ds}
\frac{d\sigma^{\pm}}{dQ^2 dy dt}=\frac{|T^{\pm}|^2}{32 (2\pi)^3
 Q^2 s^2 y}.
\ee
For the spin-average  amplitude square we find
\be
 |T^{+}|^2=  N ((2-2 y+y^2) m_J^2 + 2(1 -y) Q^2) s^2 [|B+2 m A|^2+|A|^2 |t|] I^2.
\label{t+}
\ee
Here $N$ is a known normalization factor and $I$ is the integral
 over transverse momentum of the gluon
\ba
I=\frac{1}{(m_J^2+Q^2+|t|)}
\int \frac{d^2l_\perp (l_\perp^2+\vec l_\perp \vec \Delta)}
{(l_\perp^2+\lambda^2)((\vec l_\perp+\vec \Delta)^2+\lambda^2)[l_\perp^2+\vec l_\perp \vec \Delta
+(m_J^2+Q^2+|t|)/4]}.
\ea
The term proportional to $(2-2 y+y^2) m_J^2$ in (\ref{t+}) 
represents the contribution of the virtual photon  with 
transverse polarization. The $2(1 -y) Q^2$ term describes the 
effect of longitudinal photons. It can be seen that for $Q^2 
\ll m_J^2$ the contribution of the longitudinal photon might be 
omitted.

The spin-dependent amplitude square looks like
\be
 |T^{-}|^2= N (2- y)  s |t|  [|B|^2+ m (A^\star B +A B^\star)] m_J^2 I^2.
\label{t-}
\ee
As a result, we find the following form of asymmetry: 
\begin{equation}
\lab{asy} 
A_{ll} \sim \frac{|t|}{s}\frac{[|B(t)|^2+ m (A(t)^\star B(t)+ A(t) B(t)^\star)]} {(2-2 y+ y^2)[|B(t)+2 m A(t)|^2+ |t| |A(t)|^2]}.
\end{equation}

The important property of $A_{ll}$  is that the asymmetry of vector 
meson production is equal to zero for the forward direction ($t=0$). 
The $A_{ll}$ asymmetry might be connected with the spin-dependent 
gluon distribution $\Delta G$ only for $|t|=0$. Thus, $\Delta G$ 
cannot be extracted from $A_{ll}$ in agreement with the results of  
\cite{mank}. The rapid energy dependence of asymmetry is the another 
important property of (\ref{asy}). It has been shown in \cite{gola_ll} 
that the $A_{ll}$ asymmetry in the diffractive processes is 
proportional to the fraction of the initial proton momentum $x_p$ 
carried off by the Pomeron. The mass of the produced hadron system is 
determined by $M^2_x \sim s y x_p$. For the diffractive  $J/\Psi$ 
production $M_x$ coincides with the vector meson mass and we find that 
$x_p \sim (m_J^2+Q^2+|t|)/(s y)$. As a result, the relevant $A_{ll}$ 
asymmetry decreases with growing energy. 

 The form of the $A_{ll}$ asymmetry depends on the ratio of the 
spin-flip to the non-flip  parts of the Pomeron coupling 
$\alpha_{flip}=A(t)/B(t)$ which have been found in 
\cite{gol_mod,gol_kr} to be about 0.1. The predicted asymmetry at 
HERMES energies is shown in Fig.\ 3. At HERA energy, the asymmetry 
will be negligible. Note that our results show the essential role of 
the " full block asymmetry" from (\ref{w-}) in the $A_{ll}$ asymmetry 
of the $J/\Psi$ production.

\vspace{-.1cm}
\begin{minipage}{8.5cm}
\hspace{-.5cm}
\epsfxsize=7.0cm
\epsfbox{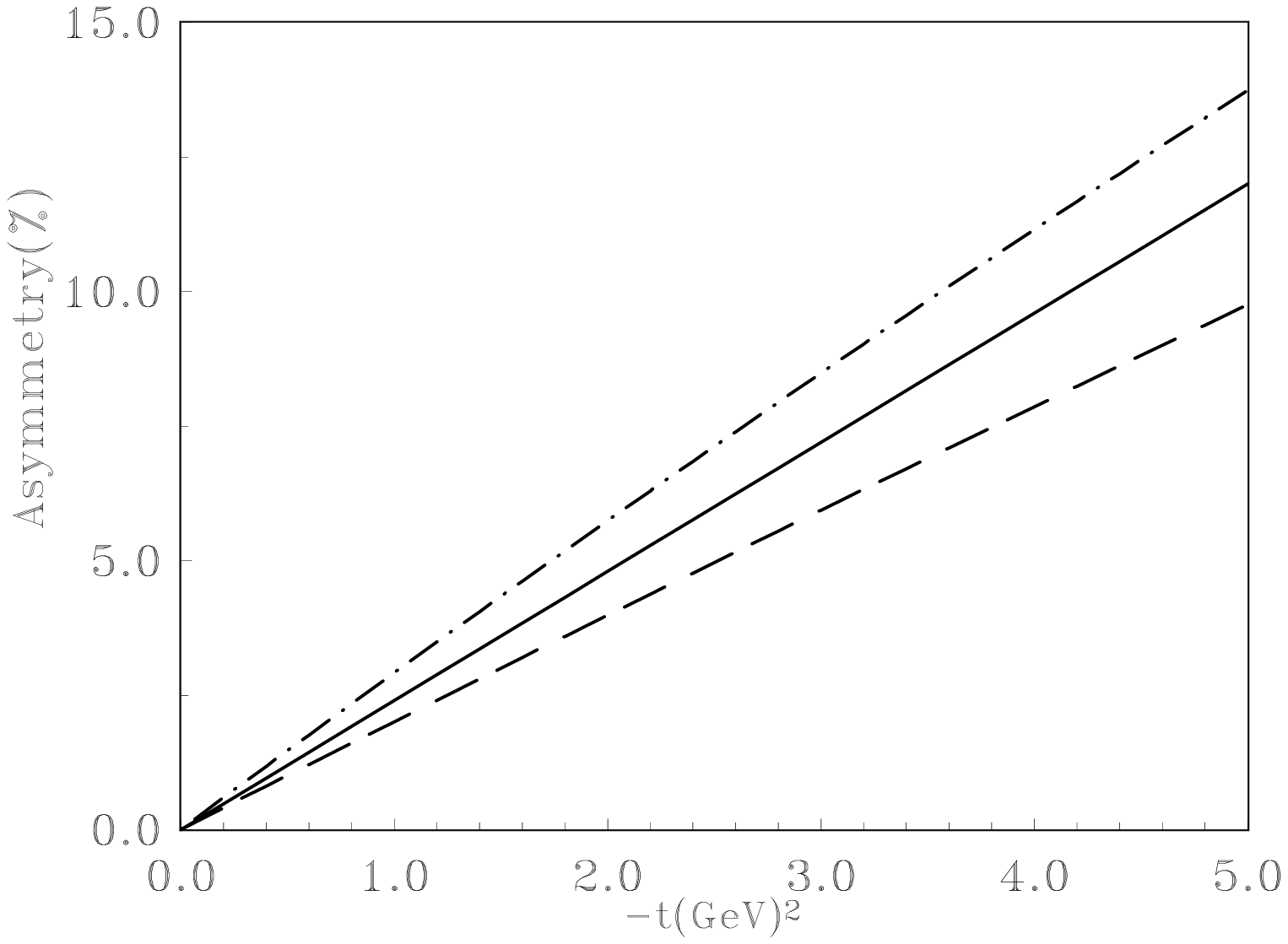}
\end{minipage}
\begin{minipage}{6.6cm}
Fig.3~ The $A_{ll}$ asymmetry of the $J/\Psi$ production  at HERMES:
solid line -for $\alpha_{flip}=0$; dot-dashed line -for 
$\alpha_{flip}=-0.1$;
dashed line -for $\alpha_{flip}=0.1$.
\end{minipage}


Let us  analyze spin effects in the diffractive $e+p \to e'+p'+Q \bar 
Q$ reaction. The difference of the polarized cross 
section  can be written in the form \cite{gola_ll}
\ba
\Delta \sigma(t)=
\frac{d^5 \sigma(^{\rightarrow} _{\Leftarrow})}{dx dy dx_p dt
dk_\perp^2}-
\frac{d^5 \sigma(^{\rightarrow} _{\Rightarrow})}{dx dy dx_p dt
dk_\perp^2}= \nonumber\\
 \frac{3(2-y)\beta_0^4 F(t)^2 [9\sum_{i}e^2_i] \alpha^2}{128
x_p^{2\alpha_{P}(t)-1} Q^2 \pi^3} \frac{A(\beta,k_\perp^2,x_p,t)}
{\sqrt{1-4k_\perp^2\beta/Q^2}(k_\perp^2+M_Q^2)^2}. \lab{dsigma}
\ea
Here $\beta_0$ is the quark--Pomeron coupling, 
$F(t)$ is the Pomeron-proton form factor and $e_i$ are the quark 
charges.
The function $A$ is determined by the trace over the quark loop.
The contribution of the standard Pomeron vertex to $A$
looks like
\be 
\lab{ad}
 A(\beta,k_\perp^2,t)=
   16 (2 (1-\beta) k_\perp^2 - |t| \beta - 2 M_Q^2 (1+\beta))
|t|.
\ee
Similar forms can be written for spin--average cross 
sections. The strong dependence of the cross sections on $\beta 
\simeq Q^2/(Q^2+M_x^2)$ and on the mass of the produced quarks 
has been found. Really, because of the negative sign in the mass 
term in (\ref{ad}) we predict positive asymmetry for light  and 
negative asymmetry for heavy quark production.

In the case of $Q \bar Q$ diffractive leptoproduction, the 
produced hadron mass is not fixed and $x_p$ is arbitrary, 
typically, of about $.05-.1$. The $A_{ll}$ asymmetry in this case 
is proportional to $x_p$ as previously and it should have a weak 
energy dependence. The predicted asymmetry for open charm 
production is about .1-.2.  Some results for the spin-dependent 
vertex and forms of the cross sections can be found in 
\cite{gola_ll}. 

To summarize, we have presented here the study of spin effects in 
diffractive processes. The discussed spin-depended form of the Pomeron 
coupling modifies spin-average and spin-dependent cross sections. The 
predicted single spin asymmetry in the elastic $pp$ scattering is about 
10-20\%. A not small value of the $A_{ll}$ asymmetry in the 
diffractive $Q \bar Q$ and $J/\Psi$ production has been found. The 
spin-dependent form of the coupling (\ref{ver}) is connected with the 
large--distance  effects in QCD. Note that the interaction of the 
gluons with the proton  can be expressed in terms of the gluon 
distribution inside the proton for zero momentum transfer \cite{rys}. 
For nonzero momentum transfer, on the other hand,  the gluon-proton 
vertex should be related to the nonforward gluon distributions. If 
so, the function $B$ might be connected with the spin-independent 
gluon distribution $G(r,x_P)$  and $A$ with the spin-dependent 
distribution $\Delta G(r,x_P)$. As a result, the study of asymmetry in 
diffractive processes might be a convenient test of the Pomeron 
coupling structure and of  nonforward spin-dependent gluon 
distributions.

The complicated form of the Pomeron vertices is determined by the 
large--distance nonperturbative effects in QCD. Thus, the important 
information on the spin structure of QCD at large distances can be 
carried out by  studying diffractive reactions in future polarized 
experiments.

\end{document}